
\input harvmac
\def\nbar{\bar{n}_A(B)}
\def\nprime{\bar{n}_A(B')}
\def\nrange{\bar{n}_A(B\rightarrow 2B)}
\def\nsbl{n_S(B,l)}
\def\Bmax{A^{1 \over {1-\gamma}}}
\def\arange{\bar{a}_A(B\rightarrow 2B)}
\def\nbl{\bar{n}_A(B,l)}

\def\nbbll{\bar{n}_A(B\rightarrow 2B,l\rightarrow 2l)}
\def\abbbb{\bar{a}_A(B\rightarrow 2B,f\sqrt{B}\rightarrow 2f\sqrt{B})}
\def\abbob{\bar{a}_A(B\rightarrow 2B,3\rightarrow fB^{\alpha})}
\def\abbll{\bar{a}_A(B\rightarrow 2B,l\rightarrow 2l)}
\def\Gi{G(A,B,l)}
\Title{\vbox{\baselineskip-10pt\hbox{HUTP-92/A002}\hbox{CTP\#2055}}}
{\vbox{\centerline{World-Sheet Geometry and Baby Universes }
\vskip2pt\centerline{in 2-D Quantum Gravity}}}
\centerline{\bf Sanjay Jain}
\smallskip\centerline{Lyman Laboratory of Physics}
\centerline{Harvard University}\centerline{Cambridge, MA 02138, USA}
\smallskip
\centerline{and}
\smallskip
\centerline{\bf Samir D. Mathur}
\smallskip\centerline{Center for Theoretical Physics}
\centerline{Massachusetts Institute of Technology}
\centerline{Cambridge, MA 02139, USA.}
\bigskip\bigskip\bigskip\bigskip
We show that the surface roughness for $c<1$ matter theories coupled to
$2D$ quantum gravity is described by a self-similar structure of
baby universes. There exist baby universes
whose neck thickness is of the order of the
ultraviolet cutoff, the largest of these having a macroscopic
area $\sim ~ A^{1 \over {1-\gamma}}$, where $A$ is the total area
and $\gamma$ the string susceptibility exponent.

\Date{04/92}

Sums over random surfaces have appeared in a number of contexts,
for example, the study of real membranes made up of molecules,
superstring theory, graph theory. In the context of membranes
they have been used to describe experimentally observed phase
transitions in which the geometric nature of the surface
changes from smooth to rough with various kinds of roughness:
crumpled, spongy, layered, etc. In this letter we are interested
in the same {\it geometric} question -- how rough is the surface
and how does one characterize its roughness --
but in the context of $c\leq 1$ conformal matter theories coupled
to 2-dimensional quantum gravity. For quantum gravity theories
this question concerns the short-distance quantum nature of
spacetime. Geometric properties of dominant surfaces in a sum
are of physical interest also for QCD and the
3-d Ising model. We hope that extracting these
properties for $c\leq 1$ models will be of use in other areas
also, particularly if they result in a geometric understanding of
the $c=1$ barrier and help to go beyond it.

The internal fractal dimension of the surface has been discussed
recently from the dynamical triangulation
\ref\AJMR{M.E. Agishtein, L. Jacobs, A.A.
Migdal, and J.L. Richardson, Mod. Phys. Lett.
A5 (1990) 965.}\ref\AM{M.E. Agishtein and A.A. Migdal,
Nucl. Phys. B350 (1991) 690, and references
therein.}\ref\KKSW{N. Kawamoto, V.A. Kazakov,
Y. Saeki, and Y. Watabiki, Kyoto and Ecole Normale preprint KUNS 1110,
HE(TH) 91/20.}
and Liouville \ref\KN{H. Kawai and M. Ninomiya,
Nucl. Phys. B336 (1990) 115.}\ref\D{F. David, Nucl. Phys. B368 (1992) 671.}
viewpoints. While there is general agreement that the typical surface
is quite rough there is yet no agreement on the value of the fractal
dimension. At present we also lack a picture of what the typical
surface looks like at various length scales and how this picture
depends upon the matter living on the surface.  From their numerical
studies of geodesic observables on the surface Agishtein and Migdal
\AM\ conclude that the typical surface for the $c=0$ system
(pure gravity) is highly branched. David \D\
provides evidence from Liouville computations
that a typical geodesic passes through many
regions where the Liouville field is highly negative (the local scale
factor is very small).

In this letter we characterize the roughness of the typical surface
in terms of the distribution of baby universes on it.
We estimate the
average number of baby universes of different areas and neck sizes,
and study the variation of these distributions with
the central charge of the matter that lives on
the surface. The results are obtained analytically by estimating the product of
entropies of the baby and parent together with the number of ways
of joining the two.

\noindent \underbar{Definition of baby universes.}
A baby universe is defined as a simply connected region of the surface
whose boundary length is much smaller than the square-root of its area.
This definition is meant to capture, in purely intrinsic terms, the intuitive
picture of a baby universe as an inflated balloon like region
of the surface attached to it by a small neck.
The thickness of a neck will be
defined as the length of the loop located at the thinnest point of
the neck. Thus, to determine the thickness we need to identify,
given a closed loop on the surface
encircling the neck of a baby universe, where
the length of this loop becomes a minimum as we slide it along the
neck. It is not enough
to have a `local' minimum, for then it is possible that a short
distance away along the neck there is another local minimum of even
smaller length, in which case the latter would be a preferable
location of the beginning of the baby universe.
It seems reasonable that in order to check for a minimum we probe
an area that is some fraction of the area of the baby universe
itself on either side of a candidate loop, and as a working definition
we choose this fraction to be a half. To be precise,
a simply connected region of area $B$ with a boundary $C$ of
length $l$ will be called a baby universe of size $B$
that begins at $C$ and has a neck thickness $l$
provided $l$ is the minimum length of all loops
to which $C$ can be continuously deformed
and which lie in any annular region of area $B/2$ on either side of $C$.
Further, $l < f\sqrt{4 \pi B}$, where $f \ll 1$,
and $B < A/2$, where $A$ is the area of the
whole surface. The last two conditions are
needed to justify the nomenclature `baby universe'.

The purpose of this definition is that we characterize a baby universe
like region of the surface by an essentially unique choice of boundary
curve $C$. (The degeneracy due to loops of exactly the same size close by
on the same neck is not expected to be significant
for our results.) Hence determining the distribution of
baby universes on a surface just involves counting loops on it
with the above mentioned properties.
A surface can have baby universes of various sizes living on it and
baby universes can grow upon other baby universes. By definition the area
of a `daughter baby universe' is smaller
than its `parent baby universe' since the
latter includes the former, but both are counted as distinct baby universes.

Baby universes with the thinnest possible neck will play
an important role. Thus, we
picture the surface as having some ultraviolet cutoff, e.g.,
a triangulated surface covered with equilateral triangles.
We will always think of a surface as defined
by such a triangulation; distinct surfaces are specified by
distinct triangulations. All lengths are in units of
the edge size of each elementary triangle, all
areas are in units of the elementary area
(a surface of area $A$ has $A$ triangles), and
matter lives on the vertices of triangles.
Thus we can imagine the thinnest neck as
having a circumference of 3; such a neck
is obtained by removing one triangle each from two closed surfaces
and joining them together at the boundary.

A baby universe whose neck thickness equals this minimum possible
circumference on the surface will be called a  `minimum neck baby universe'
(abbreviated `minbu'). The smallest possible area of a minbu is 3 (for a
3-plaquette tetrahedral `blip') and the largest minbu can have an area of
$A/2$.
\bigskip \noindent{\bf 1. Results}\hfill\break Suppose we choose a
range $B<B'<2B$ for the area of baby universes. We first mark out all baby
universes within this area range and with neck thickness   $3$. Next we
mark out baby universes (in the same size range) and neck thickness $4$,
and so on. We ask the question: What  fraction $F(f)$ of the surface is
covered by baby universes in this size range, with neck thicknesses of
order $f\sqrt{B'}$, ($0<f<1$)? If the surface was smoother at larger scales
than at smaller scales, then as we increased $B$, $F(f)$ would decrease for
small $f$ and increase for larger $f$. If on the other hand the surface was
self-similar through all scales, then $F(f)$ would be a function
independent of $B$. {\it We find that the surface is in fact self-similar},
if {\it we use in our power law ansatz for the entropy of `spheres with a
neck' an exponent indicated by the known entropy of closed surfaces of
higher genera.}

The entropy of higher genus surfaces is used in the following way. A genus one
surface can be obtained from a genus zero
surface with two equal holes by identifying
the holes. Thus with an ansatz for  the entropy of genus zero surfaces
with holes we can estimate the entropy of higher genus surfaces in terms of
the power law exponents in the ansatz. A simple argument gives
that the string susceptibility $\gamma(g)$ increases linearly with genus $g$.
Identifying the slope in the linear growth with the known value we deduce
the value of the `boundary length' exponent. This value of the exponent gives
rise to the self similar world-sheet structure mentioned above.

We now describe the results in more detail. $\nbl$ denotes the average
number of baby universes of fixed area $B$ and neck thickness $l$
on a typical closed surface of area $A$. $n_A(B_1 \rightarrow B_2,l)$
denotes the average number of baby universes of area in the range $B_1$ to
$B_2$ and fixed neck thickness $l$. The average is always taken
in the ensemble of closed surfaces of fixed area $A$. $n_A(B,3) \equiv
\nbar$ refers to minbu's.

\noindent \underbar{A. Distribution of minimum neck baby universes.}
We find that the average number of minbu's of area $B$ on a closed genus $g$
surface of area $A$ is given by
\eqn\navg{\nbar \simeq k~ A^{3 - \gamma(g)}~(A-B)^{\gamma(g) - 2}~
B^{\gamma - 2}.}
Here $\gamma(g)=\gamma + g(2-\gamma)$, where
$\gamma$ is the string susceptibility exponent, given by
$\gamma = {1 \over {12}}[c-1 -\sqrt{(25-c)(1-c)}]$ for $c\leq 1$
matter theories coupled to gravity
\ref\KPZ{V.G. Knizhnik, A.M.
Polyakov, and A.B. Zamolodchikov, Mod. Phys.
Lett. A3 (1988) 817.}\ref\DDK{F. David, Mod. Phys. Lett. A3 (1988) 1651;
J. Distler and H. Kawai, Nucl. Phys. B331 (1989) 509.}.
$k$ is a constant independent of $A$ and $B$ and of order unity for $c<1$.
(Henceforth we will not display such constants.)
As will be seen in section 2, the only assumption needed to prove \navg\
is that the partition function for closed genus zero surfaces of area $B$
is given by $Z(B) \sim e^{\mu B}~B^{\gamma - 3}$,
and the partition function for genus $g$ closed surfaces
of area $A$ is given by $Z(A) \sim e^{\mu A}~A^{\gamma(g) - 3}$.
Thus \navg\ holds whenever $A$ and $B$ are large enough for these
formulae to be valid. \hfill\break
\noindent \underbar{Corollaries:}
For $B \ll A$, \navg\ reduces to $\nbar \sim A~B^{\gamma - 2}$, in
keeping with the expectation that the local structure of a sufficiently
large surface should be independent of the genus.
The average number of minbu's of size between $B$ and $2B$ is
\eqn\navgi{\nrange = \sum_{B'=B}^{2B}\nprime \rightarrow
\int_B^{2B}dB'~\nprime \sim A~B^{\gamma-1}.}
This implies that the surface always has small minbu's,
the average number decreasing with increasing size of the minbu.
Strictly speaking \navgi\ is derived for large $B$, but if we extrapolate
it down to $B$ of order 3 (i.e., blips), it would imply
that the average number of blips is of the order $A$, i.e., a significant
fraction of
the surface is covered with blips. In the large $B$ domain $\nrange$
decreases to $O(1)$ for $B \sim A^{1 \over
{1-\gamma}}$, implying that the largest minbu to be seen on a
surface of sufficiently large area $A$ is typically of size
\eqn\bmax{B_{max}\sim \Bmax.}
Thus for $\gamma < 0$ (i.e., for $c < 1$)
 the size of the
largest minbu on the surface is always less than $O(A)$ for sufficiently
large $A$.
The average area residing in minbu's of size $B$,
i.e., the sum of the areas of all minbu's of size $B$
is $B~\nbar$, hence the average area residing in minbu's of size between
$B$ and $2B$ is given by $\arange=\int_B^{2B}dB'~B'~\nprime \sim
A~{1 \over {\gamma}}[(2B)^{\gamma}-B^{\gamma}].$
For large $B$ this decreases monotonically as $B$ increases, like $A~B^\gamma$.

Note that the $B$ dependence in \navgi\ changes character at the scale
$B \sim A_0 \equiv \exp({1 \over {|\gamma |}})$,
which is $O(1)$ at $c=0$ and infinite at
$c=1$. For $c$ close to $1$, $B^\gamma \simeq 1$ in the
domain $1 \ll B \ll A_0,$
and if \navgi\ were valid at these scales, it would imply
$\nrange \sim A/B$  and $\arange \sim A$. This would mean
that minbu's in every area range $B$ to $2B$
capture the same total area, which is $O(A)$ and independent of $B$.
\medskip
\noindent \underbar{B. Baby universes with arbitrary neck size.}
We estimate
the average number of baby universes of area $B$ and
neck size $l$ on a genus $g$ surface of area $A$ to be given by
\eqn\navgbl{\nbl \sim A^{3 - \gamma(g)}~(A-B)^{\gamma(g) -
2}~B^{\gamma -2}~l^{-(1+2\gamma)}.}
The derivation of this formula uses an ansatz (Eq. (9)) discussed in
section 2. \navgbl\ is valid only when $B$ is sufficiently large and for $l$
less  than some fraction of $\sqrt{B}$.\hfill \break
\noindent\underbar{Corollaries:}
Again for $B\ll A$, $\nbl \sim A~B^{\gamma - 2}~l^{-(1+2\gamma)}$,
independent of genus. The number of baby universes of area between $B$ and $2B$
and neck thickness between $l$ and $2l$
 is the sum $\sum_{B'=B}^{2B}~\sum_{l'=l}^{2l}~\bar{n}_A(B',l')$ which is
given by
\eqn\bbll{\nbbll \sim A~B^{\gamma-1}~l^{-2\gamma}.}
The same result (upto a numerical factor of order unity) is obtained
by including baby universes of neck size smaller than $l$ since the
sum over $l'$ is dominated by the larger necks for $\gamma < 0$.
The area carried by this set of baby universes is
$\abbll \sim A~B^{\gamma}~l^{-2\gamma}$. If we set $l$ to be a fixed fraction
$f$ of $\sqrt{B}$, we get a result that is {\it independent} of the scale $B$:
\eqn\bbbb{\abbbb \sim A~f^{-2\gamma}.}
Thus the typical surface is {\it self-similar} at sufficiently large length
scales, since, given a scale $B$, the fraction of the total area captured
by baby universes defined by that scale and $f$ (namely, those
whose area is of order $B$ and whose neck thickness
is of the order of  $f\sqrt{B}$) is independent
of the scale $B$ and depends only on $f$. \bbbb\ implies that $F(f)
\sim f^{-2\gamma}$. The $f$ dependence of this
quantity involves $\gamma$ and can be used to distinguish theories
with different central charges. As $c$ increases
it implies a drift towards narrower necks, and hence rougher surfaces.
The total area captured by baby universes defined by the scale $B$ is
of the order of the area of the whole surface $A$ for any $B$, which means that
small baby universes must live on larger ones in a self similar way.
 Note that this self similarity is a
consequence of the very specific power law $l^{-(1+2\gamma)}$ in \navgbl.

\medskip
\noindent \underbar{C. Baby universes at $c=1$.}
As $c$ approaches $1$ from below, $\gamma$ approaches zero from
below and $A_0$ diverges. Thus if \navg\ were valid at scales $1 \ll B \ll
A_0$ for $c \leq 1$, the
 baby universe distributions discussed
above for $B \ll A_0$ would be valid for the whole surface at $c=1$.
I.e., the surface is bubbly with minbu's, with minbu's
upon minbu's from the smallest to the
largest size in such a way that minbu's in {\it every} area range $B
\rightarrow 2B$
capture the {\it same} total area of the order of the
area of the whole surface, $A$, leaving little space for
baby universes on the surface with neck sizes much larger than the
minimum. This would be quite interesting since it would mean
that the picture of the surface at $c=1$ is already
visible at $c<1$ provided we look at baby universes of area less
than $A_0$. This would suggest that $c=1$ is some kind of a phase transition
point and $A_0(c)$ defines a `correlation length' on the surface that diverges
as the critical point is approached. (Note
that this divergence has the form $\exp (const./\sqrt{1-c})$
rather than  a power law in $1-c$.)

However, at $c=1$, logarithmic scaling violations
\ref\KM{V.A. Kazakov and A.A. Migdal, Nucl. Phys. B311 (1989) 171.}
are believed to modify the fixed
area partition function to $Z(A) \sim ~e^{\mu A}~A^{-1}~(A\ln A)^{2(g-1)}$
\ref\GK{See, e.g., D.J. Gross and I. Klebanov, Nucl. Phys. B334 (1990) 475.}.
That in turn modifies \navg\ to $\nbar \sim A~(B~\ln B)^{-2}$
(instead of just $A~B^{-2}$ for $\gamma = 0$).
Then minbu's at all length scales do
not capture the same area $A$, but instead $\arange \sim A~(\ln B)^{-2}$,
i.e., larger minbu's are suppressed by the logarithmic factor.
A speculation on the distribution of baby universes with larger
neck sizes is discussed at the end of the next section.

This raises the question as to whether logarithmic scaling violations are
visible even for $c<1$ theories at scales less than $A_0$. It is
possible that subleading corrections to the asymptotic formula for
$Z(A)$ at fixed genus could lead to such a behaviour (e.g., corrections
suppressed by factors of $A^{\gamma}$).\foot{We thank F. David for
suggesting this possibility.} If so, that
would still mean that $c=1$ behaviour is
captured in $c<1$ theories at scales less than $A_0(c)$.

\bigskip
\noindent {\bf 2. Proofs} \hfill \break
Let us define some notation: The fixed area partition
function is
$Z(A)\equiv\sum_S~W(S)$,
where the sum is over all closed surfaces
(distinct triangulations) of area $A$. $W(S)$ is the weight
factor to be attached to the surface $S$
which includes the integral over matter fields.
Similarly, the partition function
for surfaces with one boundary is
$Z(A,l)\equiv\sum_{S_1}~W(S)$,
where the sum is over all surfaces of area $A$ and one boundary
of length $l$. For pure gravity
$W(S)=1$, and then $Z(A)$ is just the {\it number} of closed
surfaces with area $A$, and $Z(A,l)$ the number of surfaces
having one boundary and with area $A$ and boundary length $l$.
To make the combinatoric argument
transparent we first restrict to the pure gravity case, and further,
consider only surfaces with no
handles. The generalizations to include matter and surfaces with handles
will be discussed subsequently.

If we join two surfaces of area $B$ and $A-B$
(with $B < A-B$), each having a single
boundary of length $l$, along their boundaries, we obtain a closed surface
of area $A$ with a marked loop of length $l$ partitioning it
into parts of area $B$ and $A-B$. Since the boundary has $l$
links there are generically $l$ ways of joining the
two surfaces to obtain distinct final surfaces. (If the initial
surfaces were highly symmetric, some of these $l$ ways of joining
them would not be distinct, but for large $B$ and $A-B$ this would
happen for relatively very few surfaces.) It is obvious that
any surface of area $A$ and a marked loop of length $l$ that
partitions it into sizes $B$ and $A-B$ can be represented as
a join of the aforesaid two surfaces, and for every distinct
choice of any of the original two surfaces a distinct final
marked surface is obtained. Thus the number, denoted $\Gi$, of closed surfaces
of area $A$ with a marked loop of length $l$ that partitions
the surface into parts of area $B$ and $A-B$ is given by
$\Gi \simeq l~Z(B,l)~Z(A-B,l)$.

Consider now the quantity $\nsbl$, which is defined to be the
number of ways of marking a nonintersecting loop of length $l$ on a surface $S$
such that an area $B$ is enclosed by the loop. Then by
definition, since $\Gi$ is the total number of surfaces so marked, one has
$\Gi=\sum_S~\nsbl$. From this it follows that
the average value of $\nsbl$ in the ensemble of closed
surfaces of area $A$ equals $\Gi/Z(A)$ and hence is given by
\eqn\navgii{{\langle \nsbl \rangle}_A \simeq
{1 \over {Z(A)}} ~l~Z(B,l)~Z(A-B,l).}
\noindent \underbar{Case A: $l=3$.}
In this case the region enclosed by every such loop is a minbu of size $B$
on the surface. A slight reflection will convince the reader that
since $l=3$ is the smallest loop possible, $\nsbl$
also equals the number of minbu's of area $B$ on $S$. (If $l > 3$,
$\nsbl$ in general overcounts the number of baby universes of area
$B$, since now one can possibly slide the loop along the neck
keeping the area $B$ enclosed the same, thereby obtaining
another entry in $\nsbl$ for the same baby universe. For $l=3$,
the minimum loop length, no sliding that preserves $B$ is possible.)
Thus
\eqn\navgiii{\nbar \simeq {1 \over {Z(A)}} ~3~Z(B,3)~Z(A-B,3).}

We need to estimate $Z(B,3)$, the
number of surfaces of area $B$ and one boundary of length $3$.
Since this boundary is created by removing one triangle from
a closed surface with $B+1$ triangles,
and since this triangle can be chosen
in $B+1$ ways to give, generically, a different final surface,
it follows that $Z(B,3) \sim (B+1)~Z(B+1)$. Using this in \navgiii\
and the fact that $Z(A) \sim~e^{\mu A}~A^{\gamma - 3}$
we obtain \navg\ for $g=0$.

\noindent \underbar{Case B: $l> 3$.}
{}From the set of $\nsbl$ loops on the surface we need to select the
subset that are true boundaries of baby universes in the sense described
in the introduction.
Imagine cutting the surface into two parts along such a `true' neck. We
would like to estimate the entropy of each of the two parts. The relevant
quantities are {\it not} $Z(B,l)$ and $Z(A-B,l)$, because these just count the
number of surfaces with given area and boundary length, with no reference
to the fact that the cut was along a `minimal' loop.
We need instead the quantity
$Z_1(B,l)$ which we define as follows. $Z_1(B,l)$ is
the sum over surfaces of area $B$, boundary
$l$, with the further property that this
boundary cannot be deformed along the surface
to a smaller curve if area less than $B/2$ is swept out in the deformation.
Thus $Z_1(B,l)$ counts a subset of the surfaces
included in $Z(B,l)$; it excludes
those surfaces of area $A$ and boundary length $l$ for which the boundary is
not a minimal loop in the above sense.
Our ansatz for $Z_1$ is
\eqn\ZONE{Z_1(B,l)~\sim ~e^{\mu B} B^{\gamma-2} l^{-(1+\gamma)}.}
Substituting this in the expression
$\nbl \simeq {1 \over {Z(A)}} ~l~Z_1(B,l)~Z_1(A-B,l)$ which is analogous to
\navgii\ but counts only true necks, we get \navgbl.

\noindent \underbar{Discussion of the ansatz \ZONE.}
As a justification of \ZONE\ we start from $Z_1(B,l) \sim e^{\mu A}A^{\gamma
-2}$ $l^{-(1+\alpha)}$, where the area exponent follows from the entropy $\sim
A~Z(A)$ for locating the centre of a loop on a closed surface,
and the boundary length exponent $\alpha$ is left
undetermined. This suggests a corresponding ansatz for genus zero
surfaces of area $A$ with two holes of length $l$ each:
\eqn\ZZZ{Z_1(A,l,l)~\sim~Z(A)~[A~l^{-(1+\alpha)}]^2,}
where following the  spirit of
the definition  of $Z_1(B,l)$ it is assumed
that the boundaries cannot be deformed to
smaller loops anywhere along the surface.
Each factor of $A~l^{-(1+\alpha)}$ corresponds
to having one loop on the surface.
Identifying the two holes in the $l$ possible ways
we get for the number of surfaces of genus one
\eqn\GENUSONE{Z^{g=1}(A)\sim\int_3^{\sqrt{A}}~dl~l~Z_1(A,l,l)\sim~e^{\mu
A}~A^{-1},}
which agrees with the known result. The difference between using
$Z$ and $Z_1$ is important since the former would give
genus one surfaces with an arbitrary marked loop while the
latter gives just genus one surfaces (since it marks  a
unique loop, the smallest). The above argument
also extends to give the known linear increase
of the string susceptibility exponent with
genus, $\gamma(g)=\gamma + g(2-\gamma)$.
The extra power of $A^{2-\gamma}$ for every
handle has the simple interpretation:
$A^2$ for locating two holes, $A^{-\gamma}$ for integrating over their
boundary lengths.

The same power laws in $B$ and $l$ as in \ZONE\ have appeared
in \ref\MSS{G. Moore, N. Seiberg, and M. Staudacher, Nucl. Phys.
B262 (1991) 665.}
in the context of $Z(B,l)$, and are derived by completely different methods.
Therefore one wonders about the relationship between $Z(B,l)$ and
$Z_1(B,l)$. For example, one
might ask: can the expression on the r.h.s. of \ZONE\
correspond to $Z(B,l)$?
In fact it cannot correspond to $Z(B,l)$ since it
can be seen combinatorially that $Z(B,l)$ increases
with $l$ for small $l$ while \ZONE\ decreases with $l$.
$Z(B,l)$ is expected to have an increasing factor like
$e^{\rho l}$ $(\rho > 0)$ which cannot be present
in $Z_1(B,l)$; in the latter it would lead
to the contradiction that non-overlapping
baby universes cover more than the entire area of a typical surface.
It is interesting that the two physically distinct quantities $Z$ and
$Z_1$ differ by a large non-universal factor, but their subleading
behaviour given by the power laws seems to
be the same. At large $l$ ($l \geq \sqrt{A}$),
we expect $Z_1$ to be exponentially damped by factors of
the type $e^{-{l^2 \over A}}$, like $Z$ \MSS.

The derivation of baby universe distributions presented above
generalizes easily to higher genus surfaces. The baby universe remains
a sphere with one boundary but the parent is a higher genus surface
whose entropy is appropriately modified. Accordingly the average number
of baby universes is given by
\eqn\baby{\nbl \simeq {1 \over {Z^{(g)}(A)}}
{}~l~Z_1^{(0)}(B,l)~Z_1^{(g)}(A-B,l),}
and one uses $Z^{(g)}(A) \sim e^{\mu A}~
A^{\gamma(g)-3}$, $Z_1^{(g)}(A-B,l) \sim e^{\mu(A-B)}~(A-B)^{\gamma(g)-2}$
$l^{-(1+\gamma)}$. This completes the proof of \navg\ and \navgbl\ for
pure gravity. Of course this assumes that the
total number of plaquettes is much larger than the genus.
\medskip
\noindent \underbar{Generalization of proof for unitary $c < 1$ matter.}
One might wonder if the non-local contribution to the weight factor $W(S)$
arising from the integration
over matter permits the simple product relation
$G(A,B,l)\simeq l~Z(B,l)~Z(A-B,l)$
which was essential in the above analysis. Consider the state created
at the boundary of a baby universe on a fixed surface
by integrating matter over the baby universe. Expanding
the matter integral in a complete set of states at the neck, we find that
only the Virasoro tower above the identity can contribute because the other
primaries have vanishing one-point functions on the sphere. Further,
for narrow necks ($f \ll 1$) the identity itself gives the
predominant contribution, as it is separated
by an energy gap from the next highest
state in its Virasoro tower. From this it
follows that the above factorised form
for $G(A,B,l)$ is a good estimate, and our analysis extends to
include unitary matter living on the triangulated surfaces. We are unable to
make any statement for nonunitary matter
theories where the lowest dimension operator
is not the identity, because for these theories the entropy of fixed
area surfaces is not known.
\medskip\noindent
\underbar{Speculation for c=1.} The appearance of logs in $Z(A)$ for $c=1$
implies that one must depart from the power
law ansatz \ZONE\ for $Z_1$. We discuss
here a possible modification. We ask: what
$Z_1(A,l,l)$ needs to be substituted in \GENUSONE\ to get the genus dependence
$Z^{(g)}(A) \sim ~e^{\mu A}~A^{-1}~ (A\ln A)^{2(g-1)}$ for the closed surface
partition function? It is easy to check that $Z_1^{(g)}(A,l,l) \sim
Z^{(g)}(A) [A~l^{-1} (\ln l)^{1/2}]^2$ substituted in \GENUSONE\ gives the
correct result for $Z^{(g+1)}(A)$. This suggests that for $c=1$ the analogue
of \ZONE\ might be $Z_1^{(g)}(B,l) \sim Z^{(g)}(B)~B~l^{-1}(\ln l)^{1/2}$.
Substituting this ansatz in \baby, we get
$\nbl \sim A~ (B\ln B)^{-2}$ $l^{-1}\ln l$
for $B \ll A$. This in turn implies $\abbbb
\sim  A~ (\ln B)^{-1} (1 + {\ln f \over
{\ln \sqrt{B}}})$, and $\abbob \sim A~\alpha^2(1 + {2\ln f \over
{\ln B^{\alpha}}})$. This would mean that baby universes whose neck
thicknesses are much less than $O(\sqrt{B})$ capture a significant fraction of
the total area of the surface, which is in keeping with the drift towards
narrower necks with increasing $c$ mentioned
in the previous section. We would like to emphasize,
however, that due to subtleties of subleading effects at $c=1$
which are related to the appearance of the logs, these expressions
are at the moment on a weaker footing than the $c<1$ results.
It is necessary to check our speculated ansatz for $Z_1$ by other
methods.

\bigskip
\noindent {\bf 3. Discussion}\hfill \break
We have analysed the distribution of baby
universes on randomly triangulated surfaces,
taking into account the entropies of the
baby and the parent. It would be interesting to
analyse along these lines the random `triangulations' of
3-dimensional \ref\threedim{M.E. Agishtein and A.A. Migdal,
Mod. Phys. Lett. A6 (1991) 1863; J. Ambjorn and S. Varsted, Phys. Lett.
B266 (1991) 285; D.V. Boulatov and A. Kryzwicki, Mod. Phys. Lett. A6
(1991) 3005.} and higher dimensional manifolds.

We have obtained a simple argument for the linear increase of $\gamma(g)$
with genus, given the ansatz \ZONE. This linear increase is crucial
to the existence of the double scaling limit in the corresponding string
theories. The correct slope of $\gamma(g)$ is given for the same power laws
in $Z_1(B,l)$ that give self similarity of the surface geometry. This suggests
that there could be a connection between the geometrical structure on the
world-sheet and target space properties like
the Virasoro symmetry observed in matrix models.

We find that the typical surface does have minimum-sized
necks. The surface `pinches' to the ultraviolet cutoff at these
necks.
{}From a continuum viewpoint, all surfaces with minbu's are at the boundary
of the space of surfaces. A simple `block-renormalization'
of a discretely triangulated
surface would cause the minbu's to get disconnected from the parent
leaving a puncture on both pieces.
This is likely to be a generic feature of quantum gravity theories.
One would therefore have to keep track of how the punctures evolve
under the renormalization group. Alternatively, one could possibly
use a kind of `nested'
renormalization group, wherein, upon encountering a small neck, one
first integrates over the baby universe and smoothes out the neck, and
then proceeds to integrate over the parent. We believe that the
development of the RG approach for quantum gravity theories is an
important problem, relevant for understanding observed phenomena
far above the scale of fluctuations of spacetime.

It is interesting that our formulae for the number distribution of
baby universes also arise in a semiclassical
calculation of the Liouville path integral.
The Liouville action for a spherical baby universe of area $B$ and
neck $l$ is $\sim (1-\gamma)\ln(B/l^2)$ \D.
Taking the entropy arising from different possible
placings of the baby on the parent to be\foot{It
seems that one should take the entropy to
be $\ln(A/l^2)$ instead of the $\ln A$ expected
from a Kosterlitz-Thouless type argument.
This could be because the Liouville ansatz
is to be interpreted as an effective theory,
in which, for configurations such as the
above, gravitational fluctuations smaller than the neck size
which do not destroy the essential configuration,
are already integrated over.} $\ln(A/l^2)$, one gets the contribution of
this configuration to the path integral to
be $e^{\ln(A/l^2)-(1-\gamma)\ln(B/l^2)}=AB^{\gamma-1}l^{-2\gamma}$.
We observe that this expression is the same
as our eq. \bbll\ for the average number
of baby universes of area $\sim B$ and neck
$\sim l$.
In particular, setting the free energy $\ln(A/l^2)-(1-\gamma)\ln(B/l^2)$
to be unity for $l$ equalling the cutoff
gives \bmax\ for the largest minbu, which,
as a special case implies that for $c<1$ there are
no minbu's of area $\sim A$ \D. Note that
if we assume the expression for $\nbbll$
is independent of the cutoff then self-similarity
follows immediately; $B$ and $l$ would enter
only in the combination $B/l^2$ (at least for $B \ll A$).

\noindent {\bf Acknowledgements}\hfill \break
We would like to thank M. Bershadsky, R. Brooks, R. Brower, A.A. Migdal,
and C. Vafa for discussions. We are especially grateful to F. David
for discussions and critical comments. This work is supported by
NSF grant PHY-87-14654 and the Packard Foundation.

\listrefs
\end